\RequirePackage{fix-cm}
\documentclass[smallextended]{svjour3}       
\smartqed  
\usepackage{graphicx}
\journalname{Eur. Phys. J. Plus}
\usepackage{epsfig}
\usepackage{amsfonts,amssymb}
\usepackage{color}

  \usepackage{slashed}
  \usepackage{url}
\usepackage{bm}
\usepackage{verbatim}
\usepackage{float}

\begin{document}

\title{A theorem on the Higgs sector of the Standard Model
}


\author{Marco Frasca         
}


\institute{Marco Frasca \at
              Via Erasmo Gattamelata, 3 \\
							00176 Rome (Italy)
              \email{marcofrasca@mclink.it}           
}

\date{Received: date / Accepted: date}

\maketitle

\begin{abstract}
We provide the solution of the classical theory for the Higgs sector of the Standard Model obtaining the exact Green function for the broken phase. Solving the Dyson-Schwinger equations for the Higgs field we show that the propagator coincides with that of the classical theory confirming the spectrum also at the quantum level. In this way we obtain a proof of triviality using the K\"allen-Lehman representation. This has as a consequence that higher excited states must exist for the Higgs particle, representing an internal spectrum for it. Higher excited states have exponentially smaller amplitudes and so, their production rates are significantly depressed.
\end{abstract}

\section{Introduction}

After the confirmation of the existence of the Higgs particle \cite{Aad:2012tfa,Chatrchyan:2012ufa}, it is become mandatory to improve our understanding of this sector of the Standard Model, firstly postulated by Weinberg and Salam \cite{prl_19_1264,sm_salam}, both from the experimental and theoretical sides. The scalar field that describes this particle has been put forward in the sixties \cite{Englert:1964et,Higgs:1964ia,Higgs:1964pj,Guralnik:1964eu,Higgs:1966ev,Kibble:1967sv} and has never been exploited beyond a perturbative treatment to which is supposed to be amenable. Indeed, questions like the true spectrum of such a model or the exact form of the propagators were never properly answered. Rather, it is by now a well acquired fact that this theory is plagued by triviality, that is, it cannot exist as an interacting theory. This question was recently addressed, for the four dimensional case, in \cite{Frasca:2006yx,Suslov:2010rk} for the strong coupling limit and for the full range in \cite{Jora:2015yga}. Also, this theory suffers from the hierarchy problem arising from the corrections to the mass term that would imply that the mass of the Higgs particle should be as large as the Planck mass due to perturbative corrections. This aspect of the theory is under scrutiny yet and
waiting for
experimental inputs from the just restarted LHC.

In this paper we add a further exact result to the Higgs sector of the Standard model by deriving the exact spectrum of the theory in the broken phase. The interesting point is that a finite value for the quartic coupling is enough to grant exact solutions to the classical theory and, in turn, a well definite spectrum for the quantum counterpart. This spectrum is a superimposed one on the mass of the Higgs particle and appears like substructures are
at work.
This could be explained by a higher level theory as string theory or technicolor. It appears like a mathematical property of the Higgs field and should be taken into account when studying it.

As a by-product of this theorem, we provide a mathematical technique to solve partial differential equations like that for the Green function using Lorentz invariance and Fourier transform. This result is in agreement with our conclusions in \cite{Frasca:2013gba}. Here we show how the Higgs sector in the Standard Model just implies an extended spectrum of massive excitations. We confirm these results in quantum field theory by solving the Dyson-Schwinger equations of the theory
omitting
the quadratic coupling of the gauge fields to the Higgs
field and interaction with fermion fields.
This entails a proof of triviality for the theory.
We emphasize that we limit our analysis strictly to the scalar sector of the Standard Model. This is a strong approximation as other fields like fermions and gauge bosons can change the observables we consider. But this is meaningful as a leading order understanding of the behavior of this part of the model that further corrections are not expected to change dramatically.

The paper is structured as follows. In Sec.~\ref{secII} we present the exact solutions for the Higgs sector of the Standard Model and the the Green functions of the classical theory. In Sec.~\ref{secIIb} we present the solution of the Dyson-Schwinger equations of the theory. In Sec.~\ref{secIIc} we perform an analysis of the running coupling of the theory. In Sec.~\ref{secIII} we derive the spectrum of the theory using the Feynman-Kac formula. Finally, in Sec.~\ref{secIV} we yield the conclusions.

\section{Exact solutions \label{secII}}

\subsection{Model}

We consider the Higgs sector of the Standard Model given by \cite{Quigg:2013ufa}
\begin{equation}
    {\cal L}_H=\partial_\mu\Phi^\dagger\partial^\mu\Phi-\mu^2|\Phi|^2-\lambda|\Phi|^4
\end{equation}
being
\begin{equation}
   \Phi =\left(
	 \begin{array}{c}
		 \phi^{+} \\
		 \phi^{0}
	 \end{array}
	\right)
\end{equation}
so that $|\Phi|^2=|\phi^{+}|^2+|\phi^{0}|^2$. From this one has the following equations of motion
\begin{eqnarray}
\label{eq:eom}
     \partial^2\phi^{+}&=&-\mu^2\phi^{+}-2\lambda(|\phi^{+}|^2+|\phi^{0}|^2)\phi^{+} \\
		 \partial^2\phi^{0}&=&-\mu^2\phi^{0}-2\lambda(|\phi^{+}|^2+|\phi^{0}|^2)\phi^{0}.
\end{eqnarray}
We assume $v^2/2=-\mu^2/\lambda$ for the vacuum expectation value of the theory,
as usual in the Higgs mechanism. The coupling with the gauge fields yields the Lagrangian
\begin{equation}
    {\cal L}_H={\cal D}_\mu\Phi^\dagger{\cal D}^\mu\Phi-\mu^2|\Phi|^2-\lambda|\Phi|^4
\end{equation}
being \cite{Quigg:2013ufa}
\begin{equation}
    {\cal D}_\mu = \partial_\mu+i\frac{g'}{2}{\cal A}_\mu+i\frac{g}{2}{\bm\tau}\cdot{\bm b}_\mu
\end{equation}
and the equations of motion change accordingly. The presence of fermion fields $\psi_f$ will complete the Higgs sector as
\begin{equation}
    {\cal L}_F=-\sum_fY_f[\bar\psi_{fR}\Phi^\dagger\psi_{fL}+\bar\psi_{fL}\Phi\psi_{fR}]
\end{equation}
being $Y_f$ the Yukawa couplings. This appears like a current term for the equation of the scalar field and so, it is not relevant for the determination of the spectrum. The situation is somewhat different for the gauge fields where a term depending on the square of the scalar field appears. This is depressed as the ratio of the square of the masses of the gauge bosons to the square of the vacuum expectation value of the Higgs field. Then, we will consider it relevant just perturbatively for the computation of the spectrum and so we will omit this in our analysis.
It should be noted that, adding other fields, cannot dramatically change our conclusions for the scalar sector.

\subsection{Classical solutions\label{subs:cs}}

The equations of motion (\ref{eq:eom}) admit an exact solution by introducing two different phases, U and B, and writing down
\begin{eqnarray}
\label{eq:exsols}
    \phi^{+}(x)&=&e^{i\theta_+}\varphi_{U}(x) \\
		\phi^0(x)&=&e^{i\theta_0}\varphi_{U,B}(x)
\end{eqnarray}
being for the unbroken phase, $\mu^2>0$,
\begin{equation}
   \varphi_U(x) = \Lambda\left(\frac{1}{3\lambda}\right)^\frac{1}{4}{\rm sn}\left(p\cdot x+\chi_U,-\frac{\Lambda^2\sqrt{3\lambda}}{2\mu^2+\Lambda^2\sqrt{3\lambda}}\right)
\end{equation}
provided the following dispersion relation does hold
\begin{equation}
\label{eq:drun}
    p^2=\mu^2+\frac{\Lambda^2}{2}\sqrt{3\lambda}.
\end{equation}
This solution is obtained with $\Lambda$ and $\chi_U$ being two integration constants and sn a Jacobi elliptic function having parameter $k^2=-\frac{\Lambda^2\sqrt{3\lambda}}{2\mu^2+\Lambda^2\sqrt{3\lambda}}$. These solutions hold with the condition $\mu^2>0$. We see that, already at classical level, the mass $\mu$ gets renormalized by the self-interaction with a coupling $\lambda$. It is enough that the self-coupling is finite to get such a solution. We note that this solution reduces to the known case for $\mu=0$ given in \cite{Frasca:2009bc,Frasca:2013tma}.

For the broken phase one has (the field $\phi^+$ is taken to be zero)
\begin{equation}
   \varphi_B(x) = \left(\frac{|\mu|^2}{3\lambda}\right)^\frac{1}{2}\ {\rm dn}\left(p\cdot x+\chi_B,-1\right)
\end{equation}
being dn another Jacobi elliptic function of parameter $k^2=-1$, with $\chi_B$ the phase and $\Lambda$ the energy scale, both integration constants. This holds provided the following dispersion relation holds
\begin{equation}
\label{eq:drB}
   p^2=\frac{|\mu|^2}{3}.
\end{equation}
So, our exact solution describes oscillations around the value $v$ as expected from a physical point of view. 

\subsection{Green functions}

The Green function for the theory can be obtained by the functional derivative of the equations
\begin{eqnarray}
\label{eq:eom2}
     \partial^2\phi^{+}&=&-\mu^2\phi^{+}-2\lambda(|\phi^{0}|^2+|\phi^{+}|^2)\phi^{+}+j_+ \nonumber \\
		 \partial^2\phi^{0}&=&-\mu^2\phi^{0}-2\lambda(|\phi^{0}|^2+|\phi^{+}|^2)\phi^{0}+j_0 \nonumber \\
     \partial^2\phi^{+*}&=&-\mu^2\phi^{+*}-2\lambda(|\phi^{0}|^2+|\phi^{+}|^2)\phi^{+*}+j_+^* \nonumber \\
		 \partial^2\phi^{0*}&=&-\mu^2\phi^{0*}-2\lambda|(\phi^{0}|^2+|\phi^{+}|^2)\phi^{0*}+j_0^*
\end{eqnarray}
with respect to $j_+$ and $j_0$ and $j_+^*$ and $j_0^*$. Then, the equations for the Green functions do not imply any nonlinear term. The only physical component for the Higgs field is the real component of $\phi^{0}$. We get
\begin{eqnarray}
     \partial^2\left.\frac{\delta\phi^{0}}{\delta j_0(y)}\right|_{j_0,j_{0}^*=0}&=&-\mu^2\left.\frac{\delta\phi^{0}}{\delta j_0(y)}\right|_{j_0,j_{0}^*=0}
		 -4\lambda|\phi^{0}|^2\left.\frac{\delta\phi^{0}}{\delta j_0(y)}\right|_{j_0,j_{0}^*=0} \\
		 &&-2\lambda|\phi^{+}|^2\left.\frac{\delta\phi^{0}}{\delta j_0(y)}\right|_{j_0,j_{0}^*=0}
		-2\left.\frac{\delta\phi^{0*}}{\delta j_0(y)}\right|_{j_0,j_{0}^*=0}(\phi^{0})^2+\delta^4(x-y) \nonumber \\
     \partial^2\left.\frac{\delta\phi^{0*}}{\delta j_{0}^*(y)}\right|_{j_0,j_{0}^*=0}
		 &=&-\mu^2\left.\frac{\delta\phi^{0*}}{\delta j_{0}^*(y)}\right|_{j_0,j_{0}^*=0}
		 -4\lambda|\phi^{0}|^2\left.\frac{\delta\phi^{0*}}{\delta j_{0}^*(y)}\right|_{j_0,j_{0}^*=0} \nonumber \\
		 &&-2\lambda|\phi^{+}|^2\left.\frac{\delta\phi^{0*}}{\delta j_0^*(y)}\right|_{j_0,j_{0}^*=0}
		-2\left.\frac{\delta\phi^{0}}{\delta j_{0}^*(y)}\right|_{j_0,j_{0}^*=0}(\phi^{0*})^2+\delta^4(x-y).  \nonumber
\end{eqnarray}
$\phi^0$ and $\phi^{0*}$ must to be computed at $j_0=0$ and $j_0^*=0$ reducing at the exact solutions of the preceding section. Consistently, we can assume that
\begin{equation}
      \left.\frac{\delta\phi^{0*}}{\delta j_0(y)}\right|_{j_0,j_{0}^*=0}=
			\left.\frac{\delta\phi^{0}}{\delta j_{0}^*(y)}\right|_{j_0,j_{0}^*=0}=0
\end{equation}
and then, these equations for the Green functions reduce to a single one
\begin{equation}
     \partial^2\Delta(x,y)=-\mu^2\Delta(x,y)-\lambda(6|\phi^{0}|^2+2|\phi^{+}|^2)\Delta(x,y)+\delta^4(x-y).
\end{equation}
This yields the equation for the Green function of the Higgs field in the broken phase
\begin{equation}
     \partial^2\Delta_H(x,y)-|\mu|^2\Delta_H(x,y)+6\lambda[\varphi_B(x)]^2\Delta_H(x,y)=\delta^4(x-y).
\end{equation}
Our aim is to get the spectrum of the theory from it. This can be accomplished by noting that the homogeneous equation has the following solution
\begin{equation}
     u_1(x)={\rm sn}\left(p\cdot x+\chi_0,-1\right){\rm cn}\left(p\cdot x+\chi_0,-1\right)
\end{equation}
provided the dispersion relation (\ref{eq:drB}) holds. 
As we will see below, all we need is to evaluate $\Delta_H(x,0)$ as this distribution keeps all the information on the spectrum of the theory. In order to get this propagator we rewrite the above equation as
\begin{equation}
\label{eq:GH}
     \partial_t^2\Delta_H(x,0)-|\mu^2|\Delta_H(x,0)+6\lambda[\varphi_B(x)]^2\Delta_H(x,0)=\delta^4(x)-\Delta_2\Delta_H(x,0).
\end{equation}
and iterate starting with the solution of the equation
\begin{equation}
     \partial_t^2\Delta_H^0(x,0)-|\mu^2|\Delta_H^0(x,0)+6\lambda[\varphi_B(t,0)]^2\Delta_H^0(x,0)=\delta^4(x)
\end{equation}
being
\begin{equation}
     \Delta_H^0(x,0)=\delta^3(x)G(t)
\end{equation}
and
\begin{equation}
     G(t)=\theta(t)\frac{\sqrt{3}}{|\mu|}\ u_1(t,0)=\theta(t)\frac{\sqrt{3}}{|\mu|}\ {\rm sn}\left(\frac{|\mu|}{\sqrt{3}}t+\chi_0,-1\right){\rm cn}\left(\frac{|\mu|}{\sqrt{3}}t+\chi_0,-1\right).
\end{equation}
The phase $\chi_0$ can be taken to be zero.

This can be Fourier transformed to yield
\begin{equation}
\label{eq:Gom}
     G(\omega)=\frac{\sqrt{2}\pi^3}{K^3(-1)}\sum_{n=1}^\infty n^2\frac{e^{-n\pi}}{1+e^{-2n\pi}}\frac{1}{\omega^2-m_n^2+i\epsilon}
\end{equation}
being
\begin{equation}
\label{eq:spe}
    m_n=n\frac{\pi}{K(-1)}\frac{|\mu|}{\sqrt{3}}
\end{equation}
the mass spectrum that also entails a zero mass value, the Goldstone boson. $K(-1)$ is a complete elliptic integral of the first kind. The next iterate takes the form
\begin{equation}
     \partial_t^2\Delta_H^1(x,0)-|\mu^2|\Delta_H^1(x,0)+6\lambda[\varphi_B(t,0)]^2\Delta_H^1(x,0)=\delta^4(x)-\Delta_2\Delta_H^0(x,0)=\delta^4(x)-G(t)\Delta_2\delta^3(x)
\end{equation}
remembering that should be interpreted in the sense of distribution just noting that $\int dxf(x)\delta''(x)=f''(0)$. Therefore we will get
\begin{equation}
     \Delta_H^1(x,0)=\int d^4x'\Delta_H^0(x-x',0)\delta^4(x')-\int d^4x'\Delta_H^0(x-x',0)G(t')\Delta_2\delta^3(x')
\end{equation}
that is
\begin{equation}
     \Delta_H^1(x,0)=\Delta_H^0(x,0)-\int d^4x'\Delta_H^0(x,x')G(t')\Delta_2\delta^3(x')
\end{equation}
that we can Fourier transform to give
\begin{equation}
     \Delta_H^1(p,0)=\Delta_H^0(p,0)+{\bm p}^2\Delta_H^0(p,0)G(\omega).
\end{equation}
This procedure can be iterated how far we want and we recover the fact that we are just obtaining the correction due to momentum ${\bm p}$ in the denominator of $G(t)$ producing the full propagator. This result implies that we have immediately the exact propagator for the Higgs field in the classical theory but, assuming that translation invariance is a property of the theory, we would have got it just moving from the rest frame with a boost obtaining
\begin{equation}
\label{eq:prop}
     \Delta_H(p)=\frac{\sqrt{2}\pi^3}{K^3(-1)}\sum_{n=1}^\infty n^2\frac{e^{-n\pi}}{1+e^{-2n\pi}}\frac{1}{p^2-m_n^2+i\epsilon}
\end{equation}
Thus, we have provided in this way a general mathematical technique to solve equations like (\ref{eq:GH}) reducing them to ordinary differential equations. The linearity of the partial differential equation played a relevant role. This represents the key result of our paper. We will show in the following sections the way the mass spectrum $m_n$ enters into the theory. Also, it is important to note that higher excited states appear to be exponentially depressed and so, really difficult to observe.

Given eq.(\ref{eq:prop}), the classical theory is completely solved as already shown in~\cite{Frasca:2013tma}. We note that, notwithstanding the ``wrong'' sign in the mass term, the solutions satisfy a correct dispersion relation and so, we obtain a proper spectrum for the quantum theory.

\section{Dyson-Schwinger equations \label{secIIb}}

In order to complete our argument, we solve the quantum theory of the Higgs field. This is accomplished by solving the hierarchy of Dyson-Schwinger equations for this case using the technique devised in \cite{Bender:1999ek}. We apply our analysis to the only surviving component of the field that we dub $\phi_H$. This is needed to support the conclusions of this paper.
We emphasize that, when all the contributions to the Higgs sector are considered, Dyson-Schwinger equations cannot be amenable to the treatment we present here but this provides a qualitative understanding on its behavior and has the prerogative to be an exact results at this order of approximation.

In the Dyson-Schwinger set of equations, the equation for a n-point correlation function depends on higher order correlation functions. So, one could thing that a truncation is needed. This is not the case here as we get an exact solution. In order to get such an exact solution, we put the contribution arising from higher order correlation functions at a given order to zero. Then, we check {\sl a posteriori} that this choice is a correct one. This is possible as, in the hierarchy we obtain, such functions enter in a special way that grants that can be nullified. But the check can only be executed {\sl a posteriori} as we are going to see.

The quantum equation of motion is
\begin{equation}
    \partial^2\phi_H-|\mu^2|\phi_H+2\lambda\phi_H^3=j
\end{equation}
where $j$ contains the fermion currents coupled to the Higgs field through Yukawa couplings and the currents originating from the couplings with the gauge fields. The contribution coming from the square of the Higgs field coupled to the gauge fields is neglected as it goes with the ratio between the masses of the gauge bosons and the squared vacuum expectation value of the Higgs field that is higher. So, this latter corrections should be considered by perturbation theory eventually and will be omitted in our solution of the Dyson-Schwinger equations in a first instance.

The generating functional coming from the Standard Model can be written, in a first approximation as already stated, as
\begin{equation}
    Z[j]=Z_0\int[d\phi]e^{i\int d^4x\left[\frac{1}{2}(\partial\phi_H)^2+\frac{1}{2}|\mu^2|\phi_H^2-\frac{2\lambda}{4}\phi_H^4+j\phi_H\right]}
\end{equation}
where $Z_0$ contains all other fields of the Standard Model and we are neglecting the quadratic correction arising from the gauge fields. We average on the vacuum state $|0\rangle$ and divide by $Z[j]$ yielding
\begin{equation}
\label{eq:sf1}
    \partial^2 G_1^{(j)}(x)-|\mu^2|G_1^{(j)}(x)+2\lambda\frac{\langle 0|\phi_H^3|0\rangle}{Z[j]}=j
\end{equation}
where we have defined $G_1^{(j)}(x)=\langle 0|\phi_H|0\rangle/Z[j]$, the one-point function. Then we write
\begin{equation}
    G_1^{(j)}(x)Z[j]=\langle 0|\phi|0\rangle
\end{equation}
and we take the functional derivative with respect to $j$ obtaining
\begin{equation}
    [G_1^{(j)}(x)]^2Z[j]+G_2^{(j)}(x,x)Z[j]=\langle 0|\phi^2|0\rangle
\end{equation}
and deriving once again one has
\begin{equation}
    [G_1^{(j)}(x)]^3Z[j]+3G_2^{(j)}(x,x)G_1^{(j)}(x)Z[j]+G_3^{(j)}(x,x,x)Z[j]=\langle 0|\phi^3|0\rangle.
\end{equation}
Using eq.(\ref{eq:sf1}), this becomes
\begin{equation}
\label{eq:g1}
   \partial^2 G_1^{(j)}(x)-|\mu^2|G_1^{(j)}(x)+2\lambda\left([G_1^{(j)}(x)]^3+3G_2^{(j)}(x,x)G_1^{(j)}(x)+G_3^{(j)}(x,x,x)\right)=j.
\end{equation}
Taking $j=0$, observing that the theory is invariant by translations, that is $G_2(x,y)=G_2(x-y)$, one has the first Dyson-Schwinger equation of the scalar theory
\begin{equation}
\label{eq:g10}
   \partial^2 G_1(x)-|\mu^2|G_1(x)+2\lambda\left([G_1(x)]^3+3G_2(0)G_1(x)+G_3(0,0)\right)=0.
\end{equation}
Now, we note that $G_2(0)$ is a constant and we can introduce a renormalized mass as $\mu_R^2=|\mu^2|-6\lambda G_2(0)$ and absorb it. $G_3$ enters as a constant.
This yields
\begin{equation}
\label{eq:g101}
   \partial^2 G_1(x)-\mu_R^2G_1(x)+2\lambda[G_1(x)]^3=0.
\end{equation}
This equation can be solved exactly as we show in a moment, respecting translation invariance. Before to see this, we derive the Dyson-Schwinger equation for the two-point function. We take the functional derivative of eq.(\ref{eq:g1}) to obtain
\begin{eqnarray}
   &&\partial^2G_2^{(j)}(x,y)-|\mu^2|G_2^{(j)}(x,y)+2\lambda\left(3[G_1^{(j)}(x)]^2G_2^{(j)}(x,y)+3G_2^{(j)}(x,x)G_2^{(j)}(x,y)\right. \nonumber \\
	&&\left.+3G_3^{(j)}(x,x,y)G_1^{(j)}(x)+G_4^{(j)}(x,x,x,y)\right)=\delta^4(x-y).
\end{eqnarray}
Taking $j=0$ we finally obtain
\begin{eqnarray}
\label{eq:g20}
   &&\partial^2G_2(x-y)-\mu_R^2G_2(x-y)+2\lambda\left(3[G_1(x)]^2G_2(x-y)\right. \nonumber \\
	&&\left.+3G_3(0,x-y)G_1(x)+G_4(0,0,x-y)\right)=\delta^4(x-y).
\end{eqnarray}
Again, we note that $G_3$ and $G_4$ enters in a peculiar way and we set them to be zero. We will verify that this is true below. Then,
\begin{equation}
\label{eq:g201}
   \partial^2G_2(x-y)-\mu_R^2G_2(x-y)+6\lambda[G_1(x)]^2G_2(x-y)=\delta^4(x-y).
\end{equation}
This equation, written in this way, seems to break translation invariance due to $G_1(x)$. This is so unless we choose properly the solution of the one-point function. Indeed, the solution to eq.(\ref{eq:g10}) can be written as
\begin{equation}
    G_1(x)=\left(\frac{\mu_R^2}{3\lambda}\right)^\frac{1}{2}\ {\rm dn}\left(p\cdot x+\chi_B,-1\right)
\end{equation}
being $\chi_B$ an arbitrary integration constant and provided we put $G_3(0,0)=0$,
that is a consistent choice as we will show below,
and take the momenta $p$ so that
\begin{equation}
   p^2=\frac{\mu_R^2}{3}.
\end{equation}
For eq.(\ref{eq:g20}) this is accomplished straightforwardly by taking $\chi=-p\cdot y+\chi'_B$. Then one has, making explicit the dependence on $y$ in $G_1$,
\begin{equation}
\label{eq:g200}
  \partial^2G_2(x-y)-\mu_R^2G_2(x-y)+6\lambda[G_1(x-y)]^2G_2(x-y)=\delta^4(x-y).
\end{equation}
and we get a consistent Dyson-Schwinger equation with respect to the symmetries of the theory. 
In order to get the two-point function we determine the solutions of the equation
\begin{equation}
\label{eq:y200}
   \partial^2w(x)-\mu_R^2w(x)+6\lambda[G_1(x)]^2w(x)=0.
\end{equation}
Our solution must preserve translation invariance and so \cite{Coleman:1985}
\begin{equation}
   w_1(\zeta)=\frac{dG_1(\zeta)}{d\zeta}
\end{equation}
having set $\zeta=p\cdot x+\chi_B$. The other independent solution has the form
\begin{equation}
   w_2(\zeta)=\frac{1}{4}\zeta\frac{dG_1(\zeta)}{d\zeta}+\frac{1}{2}G_1(\zeta).
\end{equation}
We just consider $w_1(\zeta)$ solution to determine the two-point function as the other breaks translation invariance. We will have
\begin{equation}
   w_1(\zeta)=\frac{dG_1(\zeta)}{d\zeta}=\left(\frac{\mu_R^2}{3\lambda}\right)^\frac{1}{2}\frac{d}{d\zeta}
	{\rm dn}(\zeta,\kappa)=\left(\frac{\mu_R^2}{3\lambda}\right)^\frac{1}{2}{\rm cn}(\zeta,\kappa){\rm sn}(\zeta,\kappa).
\end{equation}
Then, moving to the rest frame ${\bm p}=0$, the propagator takes the simple form for $t>t'$, aside for a multiplicative constant,
\begin{equation}
\label{eq:G_2}
   G_2({\bm x}-{\bm x'},t-t')=-\delta^3({\bm x}-{\bm x'})\left(\frac{3}{\mu^2_R}\right)^\frac{1}{2}\theta(t-t'){\rm cn}\left(\frac{|\mu|}{\sqrt{3}}(t-t'),-1\right)
	{\rm sn}\left(\frac{|\mu|}{\sqrt{3}}(t-t'),-1\right)
\end{equation}
to which we have to add the similar contribution for $t<t'$. From this it is very easy to obtain the two-point function \cite{Frasca:2013gba} that coincides with that given in eq.(\ref{eq:prop}).
This will solve the equation for $G_2$ provided we, consistently, will have in the following $G_3(0,x-y)=0$ and $G_4(0,0,x-y)=0$. Indeed, one has, after currents are set to zero,
\begin{eqnarray}
    &&\partial^2G_3(x-y,x-z)+2\lambda\left[6G_1(x)G_2(x-y)G_2(x-z)+3G_1^2(x)G_3(x-y,x-z)\right. \\ \nonumber
    &&+3G_2(x-z)G_3(0,x-y)+3G_2(x-y)G_3(0,x-z) \\ \nonumber
    &&\left.+3G_2(0)G_3(x-y,x-z)+3G_1(x)G_4(0,x-y,x-z)+G_5(0,0,x-y,x-z)\right]=0 \\ \nonumber
    &&\\ \nonumber
    &&\partial^2G_4(x-y,x-z,x-w)+2\lambda\left[6G_2(x-y)G_2(x-z)G_2(x-w)
    \right. \\ \nonumber
    &&+6G_1(x)G_2(x-y)G_3(x-z,x-w)+6G_1(x)G_2(x-z)G_3(x-y,x-w)\\ \nonumber
    &&+6G_1(x)G_2(x-w)G_3(x-y,x-z)+3G_1^2(x)G_4(x-y,x-z,x-w) \\ \nonumber
    &&+3G_2(x-y)G_4(0,x-z,x-w)+3G_2(x-z)G_4(0,x-y,x-w)  \\ \nonumber
    &&+3G_2(x-w)G_4(0,x-y,x-z)+3G_2(0)G_4(x-y,x-z,x-w) \\ \nonumber
    &&\left.+3G_1(x)G_5(0,x-y,x-z,x-w)+G_6(0,0,x-y,x-z,x-w)\right]=0 \\ \nonumber
    &&\vdots
\end{eqnarray}
that are solved by
\begin{equation}
   G_3(x-y,x-z)=-12\lambda\int dx_1 G_2(x-x_1)G_1(x_1-y)G_2(x_1-y)G_2(x_1-z) 
\end{equation}
and
\begin{eqnarray}
    &&G_4(x-y,x-z,x-w)= \nonumber \\
		&&-12\lambda\int dx_1 G_2(x-x_1)G_2(x_1-y)G_2(x_1-z)G_2(x_1-w) \\ \nonumber
    &&-12\lambda\int dx_1G_2(x-x_1)\left[G_1(x_1-y)G_2(x_1-y)G_3(x_1-z,x_1-w)\right. \\ \nonumber
    &&\left.+G_1(x_1)G_2(x_1-z)G_3(x_1-y,x_1-w)
    +G_1(x_1-y)G_2(x_1-w)G_3(x_1-y,x_1-z)\right].
\end{eqnarray}
and it is easy to verify that $G_4(0,0,x-y)=0$. These hold provided that
\begin{eqnarray}
   G_4(0,x-y,x-z)&=&0 \\ \nonumber 
   G_5(0,0,x-y,x-z) &=& 0 
\end{eqnarray}
and
\begin{eqnarray}
    G_5(0,x-y,x-z,x-w)&=& 0 \\ \nonumber
    G_6(0,0,x-y,x-z,x-w)&=& 0.
\end{eqnarray}
Now we are in a position to check the consistency of our choices for the first two equations of the Dyson-Schwinger hierarchy, $G_3(0,0)=0$, $G_3(0,x-y) = 0$ and $G_4(0,0,x-y)=0$.
So, one has
\begin{equation}
   G_3(0,x-z)=-12\lambda\int dx_1 G_2(x-x_1)G_1(x_1-x)G_2(x_1-x)G_2(x_1-z) 
\end{equation}
but $G_2(x-x_1)G_2(x_1-x)=0$ as we can integrate with $x>x_1$ or $x<x_1$ and not both. Similarly,
\begin{equation}
   G_3(0,0)=-12\lambda\int dx_1 G_2(x-x_1)G_1(x_1-x)G_2(x_1-x)G_2(x_1-x) 
\end{equation}
is zero too. Then,
\begin{eqnarray}
    &&G_4(0,0,x-w)=-12\lambda\int dx_1 G_2(x-x_1)G_2(x_1-x)G_2(x_1-x)G_2(x_1-w) \\ \nonumber
    &&-12\lambda\int dx_1G_2(x-x_1)\left[G_1(x_1-x)G_2(x_1-x)G_3(x_1-x,x_1-w)\right. \\ \nonumber
    &&\left.+G_1(x_1)G_2(x_1-x)G_3(x_1-x,x_1-w)
    +G_1(x_1-x)G_2(x_1-w)G_3(x_1-x,x_1-x)\right].
\end{eqnarray}
again one has $G_2(x-x_1)G_2(x_1-x)=0$ 
for the causality property of the correlation functions so, 
the condition $G_4(0,0,x-w)=0$ 
is verified. This confirms that our solutions to the first two equations of the Dyson-Schwinger hierarchy are exact. Going to higher orders we can verify also for $G_3$ and $G_4$ and so on, iterating the procedure and confirming consistency.

This concludes the analysis of the quantum equations of the theory showing how the propagator to consider, to obtain the spectrum of the theory, is exactly the classical one given in eq.(\ref{eq:prop}). Just we note that the two-point function we obtained is that of a free theory. This can be seen immediately by using the K\"allen-Lehman spectral representation that yields the exact form of the propagator in quantum field theory. One has
\begin{equation}
   G_2(x-y)=\int_0^\infty d\mu^2\rho(\mu^2)\int\frac{d^4p}{(2\pi)^4}e^{-ip\cdot (x-y)}\frac{1}{p^2-\mu^2+i\epsilon}
\end{equation}
being $\rho(\mu^2)$ the spectral density. The spectral density contains the characteristics of the spectrum of the theory and takes the general form
\begin{equation}
   \rho(\mu^2)=\sum_nZ_n\delta(\mu^2-m_n^2)+\rho_c(\mu^2)
\end{equation}
where the first term represents the excitations of the field and the last term corresponds to the interaction producing bound states. If $\rho_c$ is missing there is no interaction at all and the theory is just trivial. This is exactly our case. We see that we have
\begin{equation}
   \rho(\mu^2)=\sum_{n=0}^\infty B_n\delta(\mu^2-m_n^2)
\end{equation}
with $B_n=\frac{\sqrt{2}\pi^3}{K^3(-1)}n^2\frac{e^{-n\pi}}{1+e^{-2n\pi}}$ and $m_n$ given by eq.~(\ref{eq:spe}) with $|\mu|$ substituted by $\mu_R$. We see immediately that no bound states are present and so the theory is trivial. We have just a superimposed spectrum of a harmonic oscillator on an otherwise free particle. $B_n$ corresponds physically to the signal strength for each state and is exponentially depressed. E.g. for $n=1$ one has 0.84 to be compared with the most recent CMS data of $0.83\pm 0.21$ for WW decay \cite{Flechl:2015xxa}. Similarly, ATLAS finds $1.16_{-0.21}^{+0.24}$ \cite{Aad:2015gba}. Currently, Errors are too large yet to draw a conclusion. Finally, we note that, with a Higgs mass of 125~GeV and a spectrum given by an integer number $n$ for this mass as in eq.~(\ref{eq:spe}), the presumed new resonance seen at LHC will be exactly at $n=6$, corresponding to the expected mass of 750~GeV.

\section{Renormalization group analysis \label{secIIc}}

Comparison with results in perturbation theory are not straightforward in our case because we are expanding around non-constant solutions of the classical theory, the main reference being in this case \cite{Frasca:2013tma}. Anyhow, a renormalization group study of the theory is also possible deriving the beta function from the 4-point correlation functions. We just note that the 2-point function does not depend on the coupling $\lambda$ and so, it says nothing about the beta function.

So, we consider the 4-point function given by
\begin{eqnarray}
    &&G_4(x-y,x-z,x-w)= \nonumber \\
		&&-12\lambda\int dx_1 G_2(x-x_1)G_2(x_1-y)G_2(x_1-z)G_2(x_1-w) \\ \nonumber
    &&-12\lambda\int dx_1G_2(x-x_1)\left[G_1(x_1-y)G_2(x_1-y)G_3(x_1-z,x_1-w)\right. \\ \nonumber
    &&\left.+G_1(x_1)G_2(x_1-z)G_3(x_1-y,x_1-w)
    +G_1(x_1-y)G_2(x_1-w)G_3(x_1-y,x_1-z)\right].
\end{eqnarray}
We note that the second term is independent on $\lambda$ as it contains the 1-point function $G_1$ going like $\lambda^{-\frac{1}{2}}$. Also, $G_3$ depends on $G_1$ and brings another identical contribution canceling the $\lambda$ factor that multiplies the term. Then, this contribution is independent from the coupling. In view of the limit of large coupling, we can safely neglect such a term. This will yield
\begin{eqnarray}
    &&G_4(x-y,x-z,x-w)\approx \nonumber \\
		&&-12\lambda\int dx_1 G_2(x-x_1)G_2(x_1-y)G_2(x_1-z)G_2(x_1-w).  
\end{eqnarray}
There is no running cut-off entering this function, so the Callan-Symanzik equation reduces to the simple relation
\begin{equation}
   \beta(\lambda)\frac{\partial G_4}{\partial\lambda}+4\gamma G_4=0
\end{equation}
and the proper balancing is obtained by $\gamma=-1$ and $\beta=4\lambda$ in agreement with previous results \cite{Frasca:2006yx,Suslov:2010rk,Frasca:2013tma}. This proves that the theory is infrared trivial as lowering momenta moves the coupling toward zero as it should be expected. We note that the beta function is proportional to the spacetime dimensionality. This was already observed in \cite{Suslov:2010rk}.   

\section{Spectrum of the theory \label{secIII}}

In order to get a consistent quantum theory we have to prove that the set of solutions we obtained are indeed unique. This can be seen by introducing the new variable $\zeta=p\cdot x=p_0t-{\bm p}\cdot{\bm x}$ into the more general equation
\begin{equation}
   \partial^2\varphi(x)+\mu^2\varphi(x)+\lambda\varphi^3(x)=0
\end{equation}
yielding
\begin{equation}
   p^2\varphi''(\zeta)+\mu^2\varphi(\zeta)+\lambda\varphi^3(\zeta)=0
\end{equation}
that can be stated in the form
\begin{equation}
     \varphi''(\zeta)=-\frac{\mu^2}{p^2}\varphi(\zeta)-\frac{\lambda}{p^2}\varphi^3(\zeta). 
\end{equation}
Now, we can rescale it writing $\varphi=\xi\ \tilde\varphi$ being $\xi$ a constant. One has
\begin{equation}
     \tilde\varphi''(\zeta)=-\frac{\mu^2}{p^2}\tilde\varphi(\zeta)-\xi^2\frac{\lambda}{p^2}\tilde\varphi^3(\zeta). 
\end{equation}
This is the definition of the sn Jacobi function \cite{OLBC10} provided the parameter is given by $k^2=-\xi^2\lambda/2p^2$ and the relation of dispersion (\ref{eq:drun}) holds for a proper choice of $\xi$. Similarly, for the broken phase one obtain the differential equation for the dn Jacobi function provided the dispersion relation (\ref{eq:drB}) holds and the parameter is $k^2=-1$. So, these solutions represent unique traveling waves for the equations of the scalar field and, as such, are amenable to quantization.

In order to obtain the spectrum of the theory we use the Feynman-Kac formula \cite{ZinnJustin:2002ru}. This can be justified in the following way. Let us consider the propagator as defined by
\begin{equation}
   \langle x|e^{-itH}|y\rangle=\sum_n e^{-iE_nt}\phi_n(x)\phi_n^*(y)
\end{equation}
after the introduction of the complete set of eigenstates that diagonalizes the Hamiltonian, $H|n\rangle=E_n|n\rangle$, such that $\langle x|n\rangle=\phi_n(x)$. Then,
\begin{equation}
   \langle x|e^{-itH}|0\rangle=\sum_n e^{-iE_nt}\phi_n(x)\phi_n^*(0)
\end{equation}
and the ground state is obtained as usual after a Wick rotation $\tau\rightarrow it$ and taking the limit $\tau\rightarrow\infty$. We will get
\begin{equation}
   E_0=-\lim_{\tau\rightarrow\infty}\frac{1}{\tau}\log\left(\sum_n e^{-E_n\tau}\phi_n(x)\phi_n^*(0)\right).
\end{equation}
We see that the choice $y=0$ is irrelevant to determine the spectrum of the theory. The procedure can be iterated to compute the energy levels of the higher states.

Then, we notice that the action entering into the generating functional of the theory of the theory admits a series in the currents. One has
\begin{eqnarray}
     {\cal A}[j_0,j^*_0,j_+,j_+^*]&=&\int d^4x\left[
		    \partial\phi^{+*}\partial\phi^+
				\partial\phi^{0*}\partial\phi^0-\mu^2(|\phi^+|^2+|\phi^0|^2)-\lambda(|\phi^+|^2+|\phi^0|^2)^2\right. \nonumber \\
				&&\left.+j_+^*\phi_++\phi_+^*j_++
				j_0^*\phi_0+\phi_0^*j_0
		\right].
\end{eqnarray}
So, the Taylor series takes the form
\begin{eqnarray}
   {\cal A}[j_0,j^*_0,j_+,j_+^*]&=&{\cal A}[0,0,0,0]+ \nonumber \\
	&&\int d^4x[\left.\phi^{+*}\right|_{j_+j_+^*,j_0,j_0^*=0}j_+^*+
	\left.\phi^{+}\right|_{j_+j_+^*,j_0,j_0^*=0}j_++ \nonumber \\
	&&\left.\phi^{0*}\right|_{j_+j_+^*,j_0,j_0^*=0}j_0^*+
	\left.\phi^{0}\right|_{j_+j_+^*,j_0,j_0^*=0}j_0]+ \nonumber \\
	&&\int d^4xd^4yj_0^*(x)\Delta_H(x-y)j_0(y)+O(j_+^2)+O(j_+^3,j_0^3)
\end{eqnarray}
being $\left.\phi^{+}\right|_{j_+j_+^*,j_0,j_0^*=0}$ and $\left.\phi^{0}\right|_{j_+j_+^*,j_0,j_0^*=0}$ the exact solutions given in eq.~(\ref{eq:exsols}) and the Higgs propagator will be different depending on the phase being broken or unbroken. This represents the first few terms of the generating functional of the theory and so, we can extract the spectrum from the Higgs propagator using the Feynman-Kac formula. One has, using eq.(\ref{eq:prop}),
\begin{eqnarray}
   \Delta_H(t,{\bm p})&=&\int_{-\infty}^\infty\frac{dp_0}{2\pi}e^{-ip_0t}\frac{\sqrt{2}\pi^3}{K^3(-1)}\sum_{n=1}^\infty n^2\frac{e^{-n\pi}}{1+e^{-2n\pi}}\frac{1}{p^2-m_n^2+i\epsilon} \nonumber \\
	&=&i\frac{\sqrt{2}\pi^3}{2K^3(-1)}\sum_{n=1}^\infty n^2\frac{e^{-n\pi}}{1+e^{-2n\pi}}\frac{1}{\sqrt{{\bm p}^2+m_n^2}}e^{-i\sqrt{{\bm p}^2+m_n^2}t}
\end{eqnarray}
where we used contour integration. The application of the Feynman-Kac formula at this stage is straightforward and we can conclude that {\sl the Higgs potential in the Standard Model entails a continuous spectrum of free particles having a superimposed discrete spectrum of masses.} The argument runs similarly for the unbroken phase. This means that higher excited states of the observed particle should be expected even if these are exponentially damped and so even more difficult to be seen than the ground state.
 
\section{Conclusions \label{secIV}}

We have shown how the classical theory of Higgs sector of the Standard Model has exact solutions with the correct behavior for the dispersion relation, notwithstanding the possible ``wrong'' sign in the mass term. These solutions are propagating nonlinear waves that share the property of being unique. In this way, a quantum theory can be easily developed that, in the simplest case, displays a spectrum for the mass of the Higgs particle representing some kind of internal degrees of freedom. In the broken phase, the theory admits a classical solution describing oscillations around a constant value that is never zero as expected for the spontaneous breaking of symmetry. Green functions can be exactly computed and we provide a general technique to do so. The hierarchy of Dyson-Schwinger equations of the theory can then be solved exactly confirming the structure of the propagator and the spectrum of the theory.

This mathematical proof implies that higher excited states of the current observed Higgs particles should be expected and that the observed one is just the ground state of an extended spectrum. This internal excited states could be explained by a higher level theory like string theory or technicolor where bound states display such a behavior in a natural way. With the restart of the LHC these states should be easier to see.

In a near future we will provide a more general derivation of the behavior of the Standard Model, particularly for the gauge bosons, in presence of this general formulation of the Higgs sector.




\end{document}